\documentclass[English,superscriptaddress,prl,twocolumn]{revtex4-1}
\usepackage{graphicx} 
\usepackage[colorlinks=true,linkcolor=blue,urlcolor=blue,citecolor=blue]{hyperref}
\usepackage{xcolor}
\usepackage{soul}
\usepackage[normalem]{ulem}
\usepackage{siunitx}
\usepackage{amsmath}
\usepackage[T1]{fontenc}
\usepackage{soul}

\usepackage{textcomp} 

\begin{document}

\title{Mapping the twist angle dependence of quasi-Brillouin zones in doubly aligned graphene/BN heterostructures}   

\author{J. Vallejo Bustamante*}
\affiliation{Universit\'e Paris-Saclay, CNRS, Centre de Nanosciences et de Nanotechnologies (C2N), 91120 Palaiseau, France}

\author{V.-H. Nguyen}
\affiliation{Institute of Condensed Matter and Nanosciences, Université catholique de Louvain (UCLouvain), 1348, Louvain-la-Neuve, Belgium}

\author{L. S. Farrar}
\affiliation{Universit\'e Paris-Saclay, CNRS, Centre de Nanosciences et de Nanotechnologies (C2N), 91120 Palaiseau, France}

\author{K. Watanabe}
\affiliation{Research Center for Electronic and Optical Materials, National Institute for Materials Science, 1-1 Namiki, Tsukuba 305-0044, Japan
}
\author{T. Taniguchi}
\affiliation{Research Center for Materials Nanoarchitectonics, National Institute for Materials Science,  1-1 Namiki, Tsukuba 305-0044, Japan}

\author{D. Mailly}
\affiliation{Universit\'e Paris-Saclay, CNRS, Centre de Nanosciences et de Nanotechnologies (C2N), 91120 Palaiseau, France}

\author{J.-Ch. Charlier}
\affiliation{Institute of Condensed Matter and Nanosciences, Université catholique de Louvain (UCLouvain), 1348, Louvain-la-Neuve, Belgium}
\author{R. Ribeiro-Palau*}
\affiliation{Universit\'e Paris-Saclay, CNRS, Centre de Nanosciences et de Nanotechnologies (C2N), 91120 Palaiseau, France}

\date{\today}

\newcommand{\missing}{\textcolor{orange}}

\newcommand{\Dom}{\textcolor{blue}}
\newcommand{\Jorge}{\textcolor{orange}}
\newcommand{\Reb}{\textcolor{purple}}

\maketitle



\noindent \textbf{When monolayer graphene is crystallographically aligned to hexagonal boron nitride (BN), a moiré superlattice is formed, producing characteristic satellite Dirac peaks in the electronic band structure. Aligning a second BN layer to graphene creates two coexisting moiré patterns, which can interfere to produce periodic, quasi-periodic or non-periodic superlattices, depending on their relative alignment. Here, we investigate one of the simplest realizations of such a double-moiré structure, graphene encapsulated between two BN layers, using dynamically rotatable van der Waals heterostructures. Our setup allows \textit{in situ} control of the top BN alignment while keeping the bottom BN fixed. By systematically mapping the charge transport as a function of BN angular alignment, we identify the simultaneous signatures of the original moirés, super-moirés, and a third set of features corresponding to quasi-Brillouin zones (qBZ) formed when the system’s periodicity becomes ill-defined. Comparing our measurements with theoretical models, we provide the first experimental mapping of the qBZs as a function of angular alignment.  Our results establish a direct experimental link between moiré interference and qBZ formation, opening new avenues for engineering electronic structures in multi-aligned 2D heterostructures.}

\begin{figure*}[t]
    \centering
    \includegraphics[width=1\linewidth]{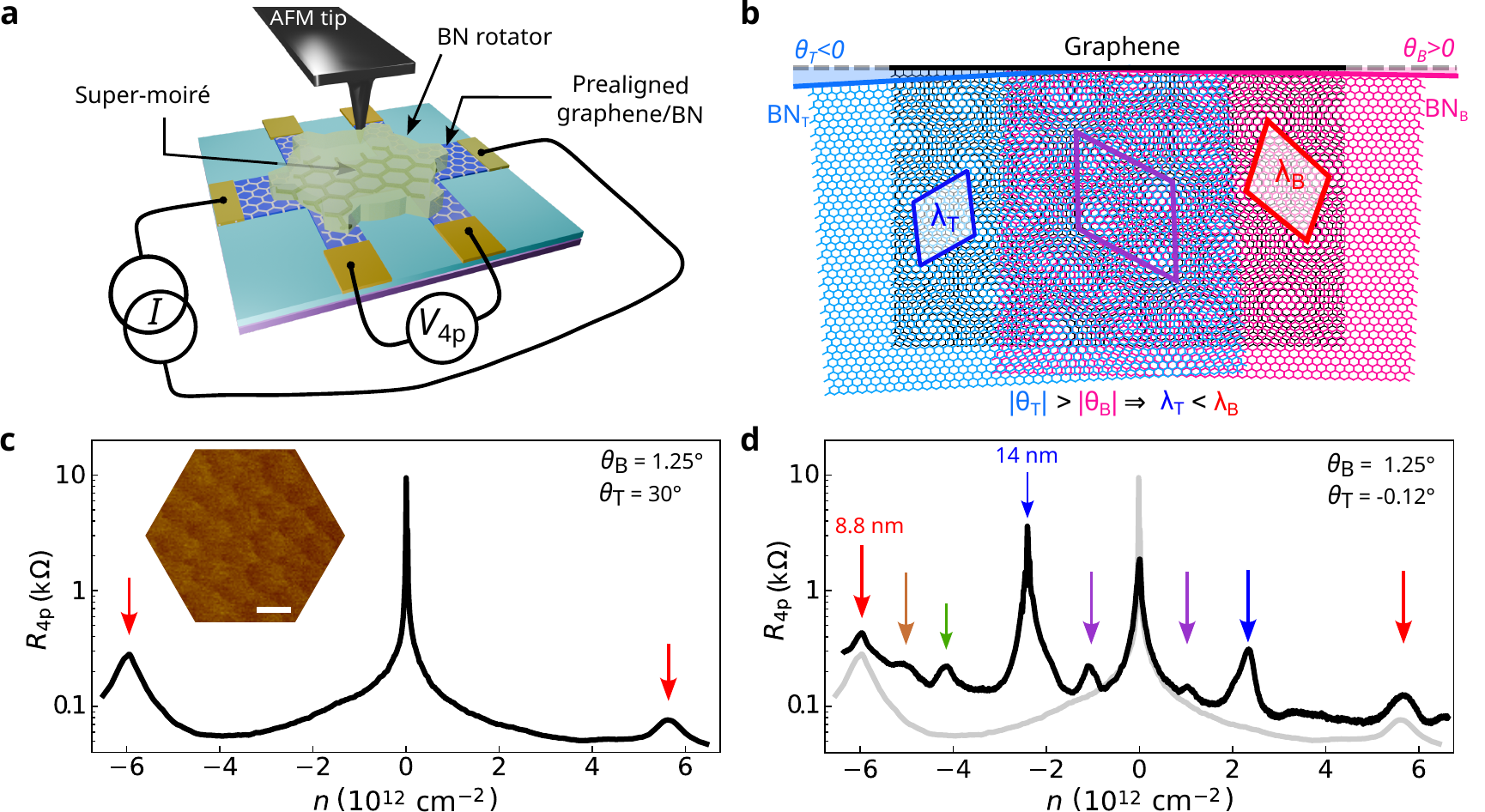}
    \caption{\textbf{Dynamically tunable double-moir\'e systems.} \textbf{a}, Schematics of a dynamically rotatable van der Waals heterostructure with pre-aligned bottom BN. \textbf{b}, Sketch of two coexisting moirés (red and blue parallelograms on each side). In the region where only the two BNs ovelap, a super-moiré lattice is indicated by a purple parallelogram.  \textbf{c}, Four-probe resistance as a function of the carrier density for $\theta_{\mathrm{B}}=1.25^{\circ}$ and $\theta_{\mathrm{T}}=30^{\circ}$. The red arrows indicate the satellite peaks from the bottom moiré. Insert: LFM scan of the sample showing the phase channel. Scale bar is 10 \unit{\nano\meter}. \textbf{d}, In black the four-probe resistance as a function of the carrier density for $\theta_{\mathrm{B}}=1.25^{\circ}$ and $\theta_{\mathrm{T}}=-0.12^{\circ}$. The gray data is plotted for comparison and corresponds to c. The red arrows indicate the satellite peaks from the bottom moiré, blue arrows the top moir\'e and the purple, green and  brown arrows indicate the peaks of the super-moir\'e. Moir\'e length for each peak is represented above the arrows. c and d charge transport measurements performed at 10 \unit{\kelvin}.}
    \label{fig:Fig1} 
\end{figure*}



The concept of a Brillouin zone, used to describe the allowed wave vectors for electrons in a periodic medium, is one of the most fundamental ones in solid state physics. {It is a powerful tool to understand the physics of a crystal.} However, when the system becomes either quasi- or non-periodic new concepts are needed to understand their physical properties \cite{oka_fractal_2021}. In recent years, the possibility to control the relative rotation of two-dimensional (2D) van der Waals (vdW) materials, and in particular graphene on hexagonal boron nitride (BN), {has} emerged as a compelling platform to investigate the effects of {a} tunable periodic potential, the so-called moir\'e superlattice \cite{hunt_massive_2013, ponomarenko_cloning_2013, dean_hofstadters_2013, ribeiro-palau_twistable_2018,inbar_quantum_2023}. The size of this periodic moir\'e potential,  can reach several orders of magnitude larger than the unit cell and causes a reconstruction of the electronic band structure \cite{yankowitz_emergence_2012}. {Superposing} two of these periodic potentials, hereafter called double-moir\'e regime, enables  to reach the quasi- and non-periodic limit and further modify the electronic band structure.

The theoretical \cite{leconte_commensurate_2020,andelkovic_double_2020} and experimental \cite{wang_composite_2019,finney_tunable_2019, sun_correlated_2021, yang_situ_2020,hu_controlled_2023,jat_higher_2024,wang_new_2019,onodera_cyclotron_2020,xie_strong_2025, lai_moire_2025} interest {in} double-moiré structures has recently increased, among others because theoretical calculations of the reconstruction of the band structure predict the appearance of flat isolated bands \cite{andelkovic_double_2020} with non-trivial topology \cite{al_ezzi_topological_2024}. However, even some of the basic features of the electronic response remain unexplained and their relation to correlated states needs to be clarified \cite{sun_correlated_2021}.

Here, we investigate the double-moiré structure made of a monolayer graphene aligned with two BN layers. We use dynamically rotatable vdW heterostructure to control and modify the alignment of the top BN, while the bottom BN is kept at a fixed alignment. By measuring the charge transport response in the region of the double-moir\'e, we are able to identify the simultaneous signatures of the original moir\'e superlattices, the super-moir\'es and the quasi-Brillouin zones (qBZ), formed when the periodicity of the system cannot be defined. Additionally, we show that the lattice relaxation plays an important role {for} the observation of the qBZ and that {at} small angles the signatures of {the} alignment can be misleading and jeopardize the understanding of the system {with} fix angular alignment.

Our samples are  fabricated following the standard dry transfer technique \cite{wang_one-dimensional_2013}, with a final step where we flip the heterostructure upside down to get an exposed graphene sample. The alignment to the bottom BN is achieved by edge alignment of the individual crystals during the stacking process \cite{yankowitz_emergence_2012}.  This is followed by micro-fabrication processes in order to contact electrically the graphene and {shape it as a} Hall bar. Once the device is processed we use the same dry transfer technique to deposit a BN rotator on top of the graphene, see Fig. \ref{fig:Fig1}a. In order to modify the crystallographic alignment between the graphene and the BN rotator we used the technique presented in \cite{ribeiro-palau_twistable_2018}. Using the tip of an atomic force microscope (AFM), {we} push the BN rotator, see Fig. \ref{fig:Fig1}a. While we change the crystallographic alignment we also monitor the four-probe electrical resistance at a finite carrier density, which allow us to know when the BN rotator is aligned (details in note 1 of the supplementary information). Low temperature measurements enable to resolve small features in the resistance and allow us to calibrate the rotation for the rest of the measured angles. In the main manuscript we present only data for sample S1, for complementary measurements in other samples see supplementary information.

When graphene is crystallographically aligned with BN, their lattice mismatch creates a  moir\'e superlattice, with a periodicity $\lambda$, that depends on: i) the smallest lattice constant, in this case graphene's lattice constant $a=0.246$ nm, ii) its mismatch to the other lattice, $\delta {\equiv} 1.78\ \%$, and of main importance in this article iii) the angle of alignment between the two crystals, $\theta$, as exemplified in Fig.\ref{fig:Fig1}b. The maximum value of the moiré superlattice created between the graphene and BN is $\lambda \simeq 14$ \unit{\nano\meter}, corresponding to a carrier density of $\pm 2.34\times 10^{12}$ \unit{\per\centi\meter\squared}. The much larger periodicity of the moir\'e superlattice, compare to the lattice constant, will create a folding of the electronic bands and therefore the opening of mini-gaps in the electronic band structure. The carrier density at which these mini-gaps are observed is given by $n=4n_{\mathrm{0}}$ \cite{hunt_massive_2013, ponomarenko_cloning_2013, dean_hofstadters_2013, ribeiro-palau_twistable_2018}, which corresponds to a doping of 4 electrons (holes) per moir\'e unit cell, whose area is $A_{\mathrm{moir}\text{\'e}}=1/n_{\mathrm{0}}=\sqrt{3}\lambda^{2}/2$, considering an hexagonal moiré unit cell.

\begin{figure*}
    \centering
    \includegraphics[width=1\linewidth]{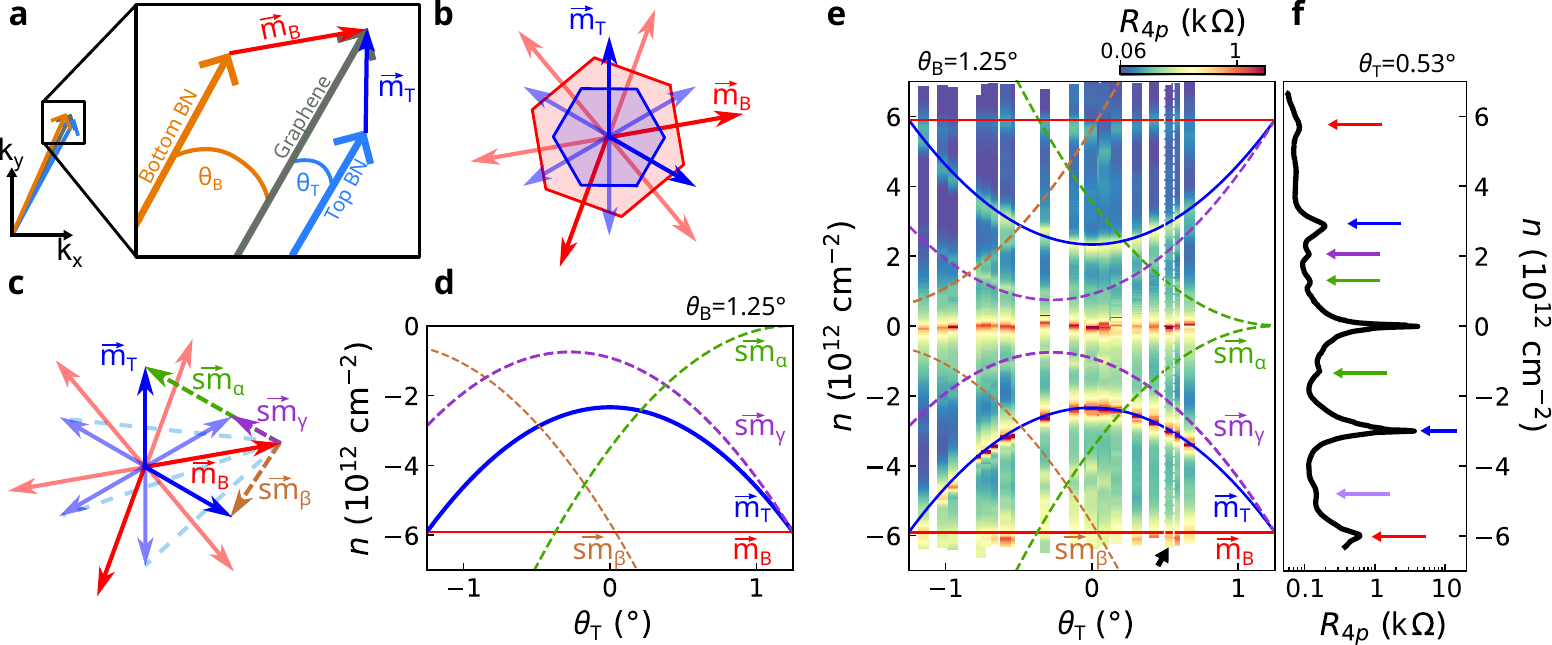}
    \caption{\textbf{Experimental results and the super-moir\'e model.} \textbf{a}, Definition of the top and bottom moiré reciprocal lattice (RL) vector as the difference between graphene and top (or bottom) BN's vectors. Due to the difference in size, only a tiny part of the original RL vectors is shown in the zoomed region. Everything else is to scale. \textbf{b}, Equivalent of a Brillouin zone (BZ) corresponding to the moirés' RL vectors. Six vectors are shown, although only two define the BZ.
    \textbf{c}, The linear combination of the two moiré RL vectors can be used to define the super-moiré lattices. These correspond to the calculated curves in d. \textbf{d}, Colormap of the angle dependence of the carrier density at full filling (four holes per moir\'e) for the equivalent BZ corresponding to the vectors in c. \textbf{e}, Four-probe resistance as a function of carrier density for different alignments of the top BN while $\theta_{\mathrm{B}}=1.25^{\circ}$, the expected trajectory of super-moir\'e peaks shown with the different dashed curves. Measurements taken at 10 K. \textbf{f}, Four-probe resistance as a function of carrier density for $\theta_{\mathrm{T}}=0.53$\unit{\degree} (black arrow in e). Each of the arrows in f points to a peak in resistance corresponding to the super-moiré model. Only the lilac arrow in this plot has no super-moir\'e correspondent.}
    \label{fig:Fig2}
\end{figure*}

In Fig. \ref{fig:Fig1}c,  we observe the  four-probe resistance as a function of the carrier density for a single-moir\'e system. This plot shows two resistance peaks (indicated with red arrows) symmetrically {spaced} around the  charge neutrality point (CNP) corresponding to the moir\'e formed between graphene and the bottom BN. From the position in carrier density of these peaks, we can deduce a moiré superlattice  of $\lambda_{\mathrm{B}}=$8.84 \unit{\nano\meter}, which is equivalent to an alignment between graphene and the bottom BN of $\theta_{\mathrm{B}}=$1.25\unit{\degree}. In this case, the top BN was misaligned with an angle $\theta_{\mathrm{T}}=$30\unit{\degree}. The angle between graphene and the bottom BN will remain fixed during all our experiments, as we cannot intentionally change it. To confirm the size of the bottom moiré, we performed lateral force microscopy (LFM) scans of the sample, see insert Fig. \ref{fig:Fig1}c, which confirms $\lambda_{\mathrm{B}}\simeq$8.8 \unit{\nano\meter}. Additionally, high temperature magneto-transport experiments show Brown-Zak oscillations, corresponding to a moiré lattice of $\lambda_{\mathrm{B}}=$8.8 \unit{\nano\meter}, see note 3 of the supplementary information.

By setting the alignment of the BN rotator to the graphene at $\theta_{\mathrm{T}}=-0.12^{\circ}$ we create two coexisting moirés modifying the charge transport response of the system. We define the rotation angle of the top BN layer as positive or negative, where negative values correspond to rotations in the direction opposite to that of the bottom BN layer, as exemplified in Fig. \ref{fig:Fig1}b. The positive/negative distinction becomes important when working with $\theta_B \neq 0$. In Fig. \ref{fig:Fig1}d we can see that in addition to the previously described satellite peaks (red arrows), we now also observe new peaks in the electrical resistance. From lower to higher densities we encounter first a pair of peaks symmetrically spaced around the CNP, their position in carrier density corresponds to a moir\'e much bigger than the allowed moir\'e between graphene and BN. These correspond to the so-called super-moir\'e \cite{wang_new_2019, wang_composite_2019, finney_tunable_2019}, a consequence of the superposition of the top and bottom moir\'es. The location in carrier density of these peaks allows for the calculation of an effective super-moiré size of $21.3$ \unit{\nano\meter}, although this size does not necessarily represent the largest periodicity of the system in real space (see note 4 of the supplementary information for more details). Following these peaks we found another pair of peaks symmetric around the CNP which correspond to the satellite peaks of the moir\'e formed with the top BN (blue arrows $\lambda = 14$ nm).  In addition to these peaks in the hole doped side we can also observed two peaks (green and brown arrows) which do not seem to have a counterpart in the electron doped side. Finally we observe the peaks corresponding to the bottom moir\'e, as previously seeing in Fig. \ref{fig:Fig1}c. As in the case in single moir\'e \cite{hunt_massive_2013, ponomarenko_cloning_2013, dean_hofstadters_2013, ribeiro-palau_twistable_2018}, here we also notice that most of the peaks in the hole side are more pronounced than their electron doped counterpart.

The super-moir\'e can be understood in reciprocal space as the superlattice resulting from a linear combination of the coexisting moirés, that form an hexagonal reciprocal lattice. In order to understand this, we use the example of a double-moir\'e with an angle $\theta_\mathrm{B}$  between the bottom BN and graphene and $\theta_\mathrm{T}$ between the top BN and graphene, see Fig. \ref{fig:Fig2}a. This angular alignment will give rise to two coexisting moir\'e superlattices  with reciprocal lattice vectors $\vec{m}_\mathrm{B}$ and $\vec{m}_\mathrm{T}$ (Fig. \ref{fig:Fig2}b). The linear combination of $\vec{m}_\mathrm{B}$ and $\vec{m}_\mathrm{T}$ results in six new super-moir\'e vectors, dashed lines in Fig. \ref{fig:Fig2}c. By keeping $\theta_{\mathrm{B}}=1.25^{\circ}$ we can model the size of each super-moir\'e assuming an hexagonal shape, and therefore calculate the position in carrier density of the resistance peaks as a function of $\theta_{\mathrm{T}}$ \cite{wang_composite_2019}, see Fig. \ref{fig:Fig2}d. For a full description of the model see \cite{wang_composite_2019} and note 4 of the supplementary information.

\begin{figure*}[t]
    \centering
    \includegraphics[width=1\linewidth]{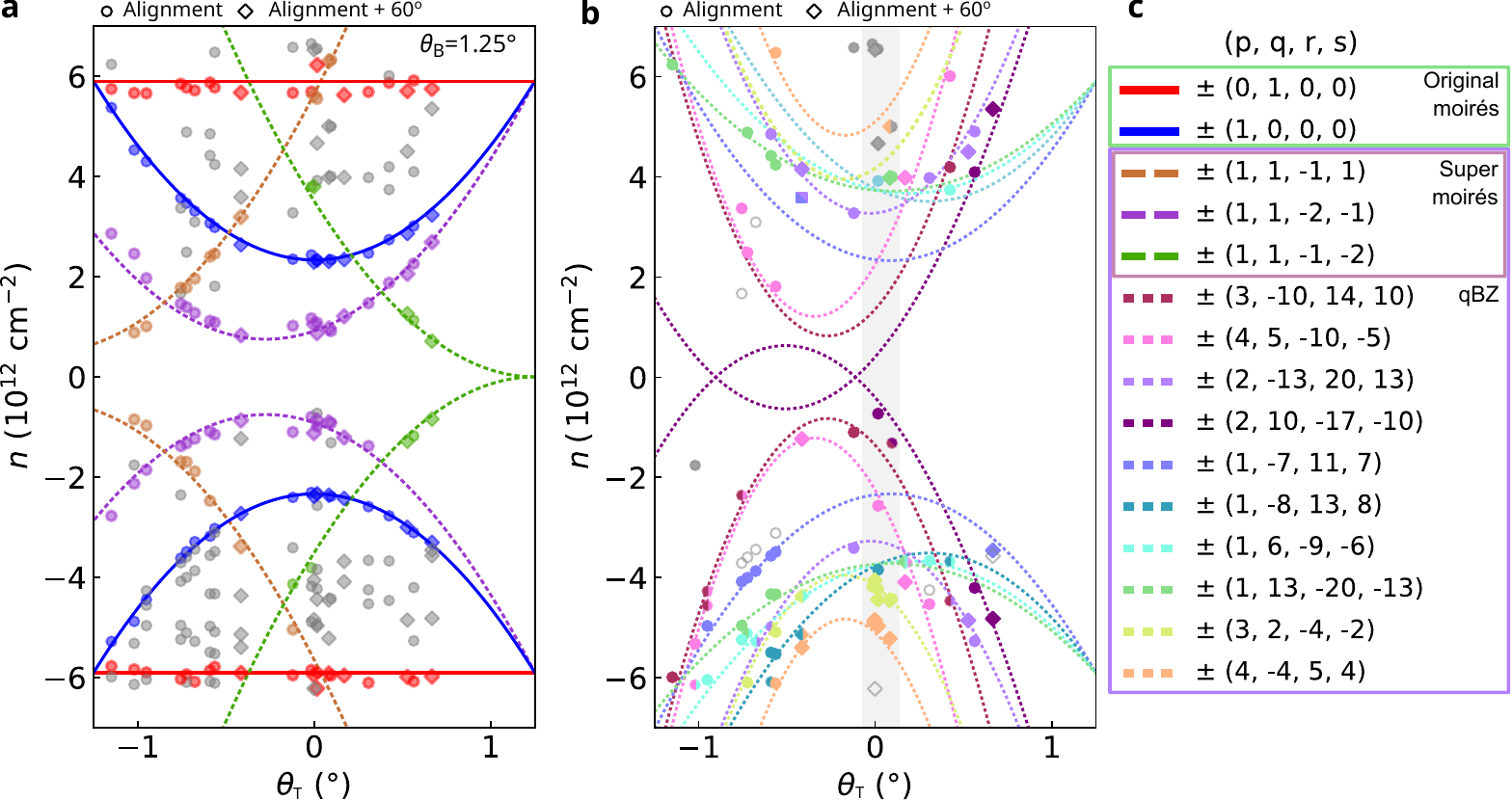}
    \caption{\textbf{Quasi-Brillouin zones model.} \textbf{a}, Summary of all the mini-gaps  observed in the resistance as a function of carrier density for for the different crystallographic alignments of the top BN. Solid lines represent the first five combinations of $(p,q,r,s)$ of the list in c, which also correspond to the super-moir\'e model. Experimental points that follow these curves have been colored for clarity. All gray points cannot be described only with this combinations. \textbf{b}, Points that cannot be explained in \textbf{a} are now fitted by using different combinations of $(p,q,r,s)$. Symbols with more than one color cannot be attributed to only one parabola. White filled symbols cannot be described by any parabola but given their proximity to another peak are believe to be angle inhomogeneities. Gray symbols cannot be explained by the model, with the used combinations of integers. \textbf{c}, Values of $(p,q,r,s)$ for the original moirés and the quasi-Brillouin zones.}
    \label{fig:Fig3}
\end{figure*}

Figure \ref{fig:Fig2}d shows the angle dependence of the position in carrier density for the full filling of each one of the super-moir\'e bands (dashed lines) \cite{wang_composite_2019} and for the two original moir\'e superlattices (solid lines), in the carrier density range accessible in our experimental setup. We show only the calculation for the hole doped side but as expected this model is electron-hole symmetric. In this figure we see a horizontal line corresponding to the fixed bottom BN alignment (red solid line). We can also identify the parabola describing the position of the satellite peak consequence of the moir\'e between the graphene and top BN (blue solid line). The three dashed curves (brown, purple and green) represent the angle dependence for the full filling of each super-moir\'e, for  angles in the range [$-1.25^\circ$, $1.25^\circ$]. It is important to highlight that the super-moir\'e described by the green dashed line represents the moir\'e superlattice between the two BN layers and for that reason its position in carrier density can reach zero, representing an infinite moir\'e, only reachable in homostructures (when the two lattices forming the moiré are identical). In Fig. \ref{fig:Fig2}e we compare directly the super-moir\'e model to our experimental results. We show a color-map of the resistance as a function of the carrier density and top BN alignment for twenty-two different crystallographic alignments on the same sample. The numerical results of the super-moir\'e model are superposed to the data to highlight the visible sets of peaks. For instance, we can see the peaks coming from the bottom alignment at a fixed position at $\pm 6\times10^{12}$  \unit{\per\centi\meter\squared}, coinciding with the red line. We can also notice the trajectory of the peaks coming from the top alignment, which coincides with the blue curve. Less prominent but still visible we are able to identify three, out of the six, super-moirés given by the model described above \cite{wang_composite_2019}. The other three super-moir\'e exist at carrier densities outside our experimental reach, see note 4 of the supplementary information. 

However, using this model we cannot explain all the features observed in our charge transport measurements. As an example, the resistance plot in Fig. \ref{fig:Fig2}f,  shows the four-probe resistance for a combination of $\theta_\mathrm{B} = 1.25^{\circ}$ and $\theta_\mathrm{T} = 0.53^{\circ}$. In this curve, besides the  resistance peaks that come from the bottom moir\'e (red arrows), the top moir\'e (blue arrows) and the super-moirés (purple and green arrows) we can also see another one (lilac arrow) in the hole doped side. These cannot be matched with any of the super-moir\'e gaps. The same occurs for all the other alignments, where we observe peaks in the resistance that cannot be associated with super-moirés. These are particularly numerous for carrier densities higher in magnitude to those that corresponding to the satellite peak of the top BN. Previous works have suggested that these features could be related to higher-order moir\'e periodicity or super-moir\'e patterns between further zone edges \cite{wang_composite_2019}. However, most of the peaks do not cluster around trajectories that are integer multiples of the moirés or super-moirés in Figs. \ref{fig:Fig2}d,e, discarding this possibility.

\begin{figure*}[t]
    \centering
    \includegraphics[width=1\linewidth]{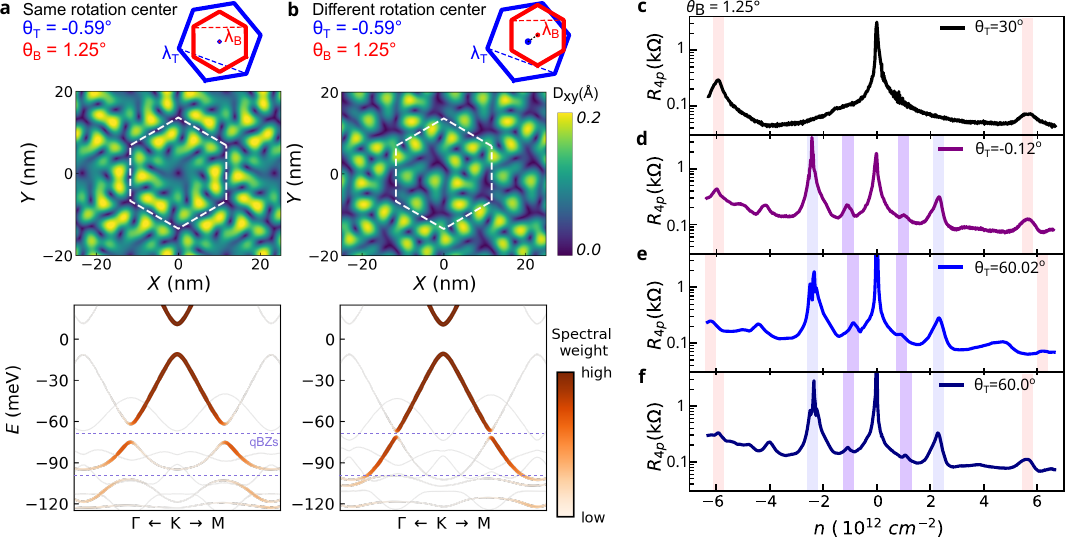}
    \caption{ \textbf{Effects of the atomic relaxation in double-moir\'e region}. \textbf{a} and \textbf{b}, (top) in plane atomic displacement, $D_{\mathrm{xy}}$, for the same combination of angular alignments $\theta_{\mathrm{T}}=-0.59^{\circ}$ and $\theta_{\mathrm{B}}=1.25^{\circ}$ with different rotation centers. Black hexagonal structure shows the superlattice unit cell. The band structure and respective spectral weight is shown in the bottom of the figure. Full energy range can be seen in Fig. S13. \textbf{c}-\textbf{f}, four probe resistance as a function of the carrier density for the same bottom alignment but four different top alignments. Color bars represent the main superlattice mini-gaps.}
    \label{fig:Fig4}
\end{figure*}

In the range where the two moirés coexist, additionally to the primary moir\'es, the relation between them will create periodic and quasi-periodic superlattices and non-periodic structures, that even in the non-periodic case, can show their signatures in the form of mini-gaps in the band structure. In order to clarify the effects of these additional superlattices in our experiments we will follow the model developed in \cite{oka_fractal_2021}. In this model the energy gaps that result from the different superlattices in the system appear at densities giving by $n = 4 A_w/(2\pi)^2$, where the area $A_w$ (in reciprocal space) is characterized by a set of integers $(p,q,r,s)$:
\begin{equation}
\label{eq:Aw}
    A_w(p,q,r,s) = pA_1 + qA_2 + rA_3 + sA_4,
\end{equation}
with
\begin{equation}
\begin{matrix}
A_1 = (\vec{m}_{\mathrm{T1}} \times \vec{m}_{\mathrm{T2}})_z, & & & & & A_2 = (\vec{m}_{\mathrm{B1}} \times \vec{m}_{\mathrm{B2}})_z,\\
A_3 = (\vec{m}_{\mathrm{T1}} \times \vec{m}_{\mathrm{B1}})_z, & & & & & A_4 = (\vec{m}_{\mathrm{T1}} \times\vec{m}_{\mathrm{B2}})_z,
\end{matrix}
\label{areas}
\end{equation}
\noindent where $\vec{m}_{\mathrm{T(B)1(2)}}$ are the reciprocal lattice vectors of the top ($\mathrm{T}$) and bottom ($\mathrm{B}$) moiré patterns. The term $(...)_z$ represents the z-component perpendicular to the plane, and it can be negative depending on the relative angles between the two vectors. $A_1$ and $A_2$ are the Brillouin-zone areas of the individual top and bottom moir\'e patterns, respectively. $A_3$ and $A_4$ are cross terms which combine the reciprocal vectors of the different moir\'e patterns. Fig. \ref{fig:Fig3}a shows the position in carrier density for all the peaks observed in the four-probe resistance as a function of the angular alignment with the top BN. Here we have represented in color: i) top moir\'e ($A_1$ - in blue), ii) bottom moir\'e ($A_2$ - in red), since it is a fixed angle the carrier density at which it is expected does not change with the alignment of the top BN, and iii) three sets of combinations of the areas $A_1$, $A_2$, $A_3$ and $A_4$, (brown, purple and green dashed lines), which are equivalent to the super-moir\'e model described before. We can see, in a clearer way, that there is still a large number of mini-gaps (gray points) that cannot be explained only by the super-moiré model.

In Fig. \ref{fig:Fig3}b we plotted the position in carrier density and angle for all the mini-gaps that cannot be associated to the super-moiré model along with the parabolas obtained by extending to a larger number of combinations of $(p,q,r,s)$. These represent areas bounded by the Bragg planes of combinations of the reciprocal vectors of the original moirés. In other words, these are polygons enclosed by multiple Bragg planes of different moirés and super-moirés and reflect the existence of quasi-Brillouin Zones \cite{oka_fractal_2021}. The integer numbers of the sets $(p,q,r,s)$ have in principle no restrictions, however according to \cite{koshino_topological_2022} the $(p,q,r,s)$ numbers can be directly related to the second Chern numbers (2CN) in the generalized version of quantum Hall effect in 4 dimensions (4DQHE), so for the 2CNs to be integer too, the sum $r+s$ must be multiple of 3, following the $120^\circ$ symmetry of the system \cite{koshino_topological_2022}. 
The 2CN are topological invariants that describe the quantized Hall response of a 4D system in the QH regime, as it has been described in \cite{kraus_four-dimensional_2013}, where it was shown that 4DQHE can be mapped into a 2D quasiperiodic crystal. Although we provide the equivalence between the four integers $(p,q,r,s)$ and the six corresponding 2CNs, in supplementary note number 6, no direct evidence of 2CN (or quantization as described in 4DQHE) has been observed in these samples at the lowest temperature. 

The mini-gaps shown in Fig. \ref{fig:Fig3}b are not expected to be observed in the full range of angular alignment \cite{oka_fractal_2021}, however we have also noticed that they are not always present inside the expected range. We can attribute this to two main factors: first, the quasiperiodic nature of the system makes the features observed in the resistance plots extremely sensitive to small variations of angle. 
These angle inhomogeneities are important at the super-moiré scale and will not be noticeable in the single moiré gaps (see Fig. S8 in the supplementary information for an example). 
Second, the very complex relaxation pattern generated in the double-moiré regime is highly sensitive to the rotation axis and translations of the layers, causing the opening and closing of different energy gaps. To illustrate this effect, we numerically simulate a BN/graphene/BN system (Figs. \ref{fig:Fig4}a,b) with identical top and bottom rotation angles but different rotation centers, corresponding to a translation of one BN layer relative to the others. Since the target angles in our experiments are generally incommensurate (i.e., the two moirés have periodicity ratios that are not rational), we use commensurate approximants that, within a chosen tolerance, accurately capture the system’s physical properties. Further details on the simulations and commensurate approximants are provided in Note 5 of the Supplementary Information.
In the top part of Figs. \ref{fig:Fig4}a and b we can see that the atomic relaxation, given by the in plane displacement of atoms $D_{\mathrm{xy}}$, presents two different patterns. The associated electronic band structure and its spectral weight (bottom part of Figs. \ref{fig:Fig4}a and b) shows that the mini-gaps of the system are impacted by the translation. However for this to be valid, the translation of the rotation axis must be comparable to the size of the commensurate approximant. Given the difference in scale of the simulated commensurate approximant ($\sim 10^1$ nm) and the size of our system ($\sim10^3$ nm), and the fact that we cannot control the rotation axis, a combination of both effects is likely to be present. We highlight that these effects do not create random qBZ mini-gaps, it just change the amplitude of the mini-gaps. We demonstrate this experimentally in the gray-shaded region of Fig. \ref{fig:Fig3}b. Near full alignment of the top BN layer ($\theta_\mathrm{T}=0$), the measurements—each taken at slightly different, non-consecutive angles, including those near $60^\circ$ alignment—consistently show gaps corresponding to the same qBZ.

The importance of the atomic relaxation (see also Fig. S7) can also be seen experimentally when paying close attention to the mini-gap corresponding to the bottom BN alignment (pink areas in Fig. \ref{fig:Fig4}c-f) for different $\theta_T$ close to alignment. For reference, the Fig. \ref{fig:Fig4}c shows the case when the top BN is misaligned. In Figs. \ref{fig:Fig4}d-f we observe that the position of both electron and hole satellite peaks appear shifted in some cases (Fig. \ref{fig:Fig4}e). This shift can be associated with a different rotation axis and therefore atomic relaxation. This will modify the satellite gap by either decreasing its amplitude it or moving it to higher energies.
We cannot rule out that by pushing the BN rotator, some strain develops on the system, which can produce a small variation on the bottom BN angle. However, the position of the super-moiré peak (purple area) shifting to lower carrier density instead of the expected higher density, makes this scenario unlikely (see Fig. S5).

This clear shift of the position in carrier density of the resistance peak related to the bottom BN alignment might be crucial when determining the original angles of a double-moiré system. Fig. \ref{fig:Fig4}e also suggests for certain atomic relaxations that the most prominent peaks are not the ones coming from the original moirés and using their position in carrier density to try to understand the electronic transport response might lead to errors. In this case, it would be necessary to rely on additional techniques to determine the moiré angles, such as magneto-transport measurements at high temperature, where Brown-Zak oscillations are visible. 

\textbf{Conclusions:} We have investigated a double-moiré structures using dynamically rotatable vdW heterostructure to control and modify the alignment of the top BN, while the bottom BN is kept at a fixed alignment. By mapping the electron transport response of a double-aligned system as a function of the twist angle we are able to identify the simultaneous signatures of the original moiré superlattices, the super-moirés and to map the qBZ. The later are formed when the periodicity of the system cannot be defined. Additionally, we have shown that the lattice relaxation plays an important role in the interpretation of the experimental data and that for small angles the signatures of alignment can be misleading and jeopardize the understanding of the system in fix-angle devices. Furthermore, the study of these quasiperiodic moiré structures might provide a promising platform to realize exotic phenomena such as the observation of the second Chern numbers, which can help understand the interplay between correlations and dimensionality.


\bibliography{references_RR}

\section*{Acknowledgments}

The authors acknowledge discussions with Adolfo Grushin, Cory Dean, Mikito Koshino and Naoto Nakatsuji. This work was done within the C2N micro nanotechnologies platforms and partly supported by the RENATECH network and the General Council of Essonne. This work was supported by: ERC starting grant N$^{\circ}$ 853282 - TWISTRONICS (R.R-P.), the DIM-SIRTEQ project TOPO2D, the DIM QuanTIP project Q-MAG and IQUPS. R.R.-P. and J.-C.Ch. acknowledge the Flag-Era JTC project TATTOOS (N$^{\circ}$ R.8010.19) and the Pathfinder project ``FLATS'' N$^{\circ}$ 101099139.  V.-H.N and J.-C.C. acknowledge financial support from the F.R.S.-FNRS through the research project “MOIRÉ” (N° T.029.22F) and the EOS project “CONNECT” (N° 40007563), from the Fédération Wallonie-Bruxelles through ARC project “DREAMS” (N° 21/26-116), and from the Pathfinder project “FLATS” (N° 101099139). Computational resources were provided by the supercomputing facilities of UCLouvain (CISM) and the Consortium des Equipements de Calcul Intensif en Fédération Wallonie Bruxelles (CÉCI) funded by the Fonds de la Recherche Scientifique de Belgique (F.R.S.-FNRS) under the convention N° 2.5020.11. K.W. and T.T. acknowledge support from the JSPS KAKENHI (Grant Numbers 21H05233 and 23H02052) and World Premier International Research Center Initiative (WPI), MEXT, Japan.

 \section*{Author Contributions Statement}
 
R.R.-P., D.M. and J.V.B. designed the experiment. J.V.B. and L.S.F. fabricated the devices. J.V.B. performed the electron transport experiments and analyzed the data.  T.T. and K.W. grew the crystals of hexagonal boron nitride. V-H.N and J.C.Ch. performed the numerical simulations. All authors participated to writing the paper. 

\section*{Competing Interest Statement}

The authors declare no competing interests.

\end{document}


\title{Mapping the twist angle dependence of quasi-Brillouin zones in doubly aligned graphene/BN heterostructures\\
Supplementary information}   

\author{J. Vallejo Bustamante*}
\affiliation{Universit\'e Paris-Saclay, CNRS, Centre de Nanosciences et de Nanotechnologies (C2N), 91120 Palaiseau, France}

\author{V.-H. Nguyen}
\affiliation{Institute of Condensed Matter and Nanosciences, Université catholique de Louvain (UCLouvain), 1348, Louvain-la-Neuve, Belgium}

\author{L. S. Farrar}
\affiliation{Universit\'e Paris-Saclay, CNRS, Centre de Nanosciences et de Nanotechnologies (C2N), 91120 Palaiseau, France}

\author{K. Watanabe}
\affiliation{Research Center for Electronic and Optical Materials, National Institute for Materials Science, 1-1 Namiki, Tsukuba 305-0044, Japan
}
\author{T. Taniguchi}
\affiliation{Research Center for Materials Nanoarchitectonics, National Institute for Materials Science,  1-1 Namiki, Tsukuba 305-0044, Japan}

\author{D. Mailly}
\affiliation{Universit\'e Paris-Saclay, CNRS, Centre de Nanosciences et de Nanotechnologies (C2N), 91120 Palaiseau, France}

\author{J.-Ch. Charlier}
\affiliation{Institute of Condensed Matter and Nanosciences, Université catholique de Louvain (UCLouvain), 1348, Louvain-la-Neuve, Belgium}

\author{R. Ribeiro-Palau*}
\affiliation{Universit\'e Paris-Saclay, CNRS, Centre de Nanosciences et de Nanotechnologies (C2N), 91120 Palaiseau, France}

\date{\today}

\newcommand{\missing}{\textcolor{orange}}

\newcommand{\Dom}{\textcolor{blue}}
\newcommand{\Jorge}{\textcolor{orange}}
\newcommand{\Reb}{\textcolor{purple}}
\newcommand{\JC}{\textcolor{red}}
\newcommand{\VietHung}{\textcolor{red}}

\maketitle

\section{Note 1: Measurements description}
\begin{figure}[h]
    \centering
    \includegraphics[width=0.6\linewidth]{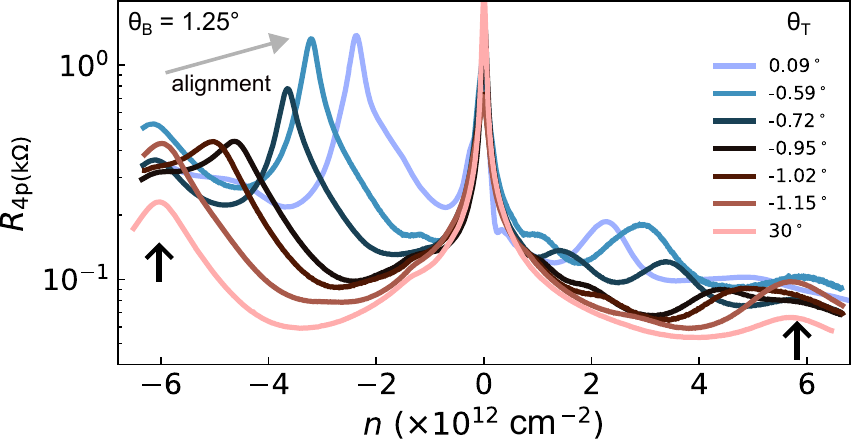}
    \caption{\textbf{Room temperature measurements}. Four-probe resistance measurement as a function of carrier density (calculated from the gate voltage) for different alignments of the top rotator, measurement taken at room temperature. Black arrows indicate the resistance peak due to the alignment with the bottom BN.}
    \label{fig:room}
\end{figure}

The changing of the angle is done at room temperature, with a current bias of 100 \unit{\nano\ampere}, in a custom sample holder inserted in our Atomic Force Microscope (AFM) \cite{ribeiro-palau_twistable_2018}. We use the tapping mode of the AFM to obtain detailed images of the rotator and the graphene below and then, it is switched to contact mode to push the rotator taking care of not passing over the graphene part. The alignment of the sample is detected electrically by measuring the four probe resistance at a certain gate voltage as a function of time in a continuous way. When the rotator gets to crystallographic alignment with the BN the electrical signal increases. The fine tuning of the angle is made by measuring the four probe resistance as a function of gate voltage for different angular alignments, as shown in fig.\ref{fig:room}. Once the alignment is set, the sample is cooled down. 

\section{Note 2: Samples S1 and S2}
\begin{figure}[h]
    \centering
    \includegraphics[width=0.43\linewidth]{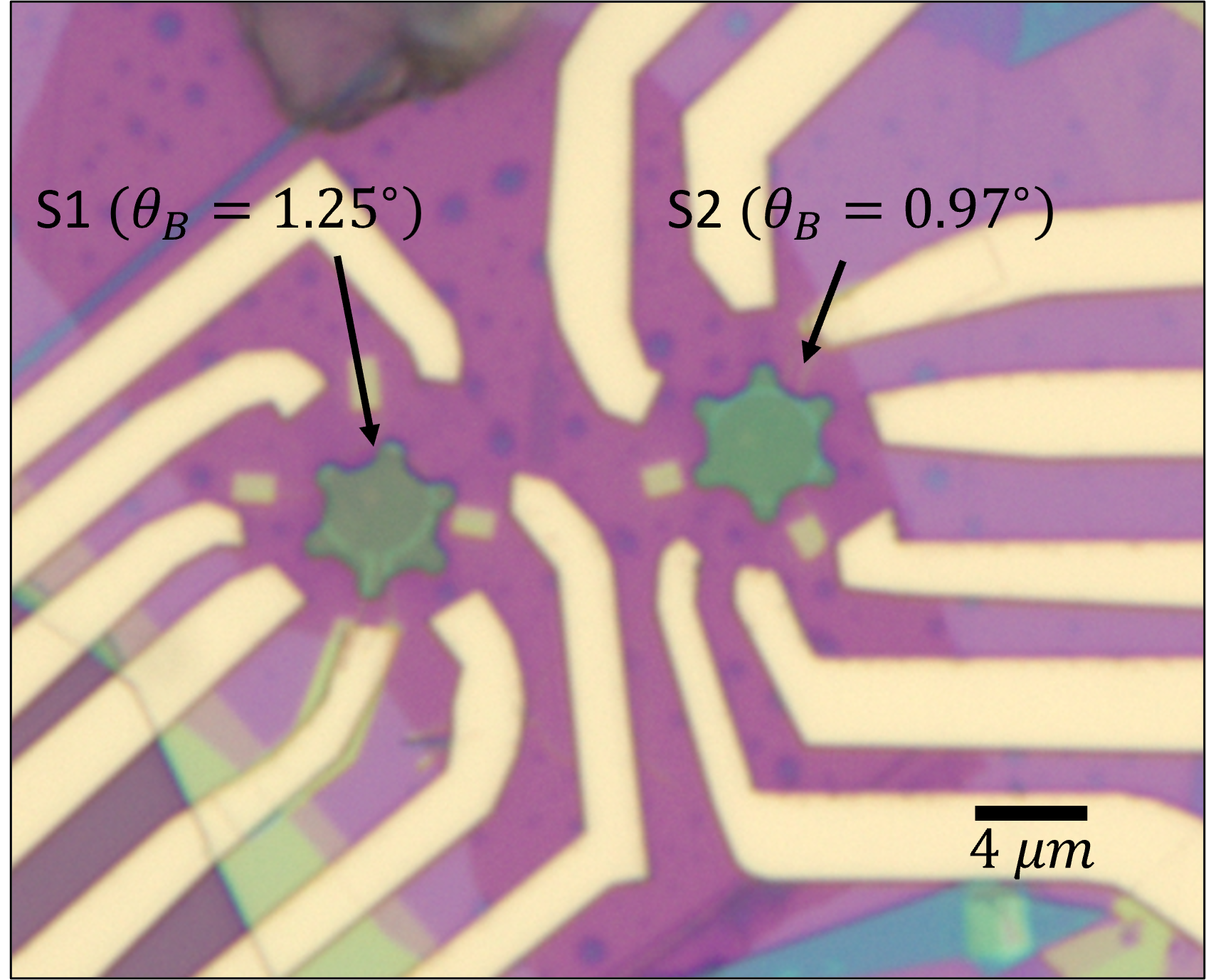}
    \caption{\textbf{Samples S1 and S2}. Dynamically rotatable van der Waals heterostructure with double aligned structure. The bottom angles are indicated in the figure.}
    \label{fig:samples}
\end{figure}

The heterostructures used in this manuscript are built as described in the main text. A total of three samples where measured: S1 (JV3B), S2 (JV3A) and S3 (LSF-M1). S1 and S2 can be seen in Fig. \ref{fig:samples}. A fold in the graphene at the moment of the pick-up allows us to make two samples with intentionally different bottom alignments on the same substrate. S3 has a bottom BN alignment of $\theta_{\mathrm{B}}=0.7^{\circ}$.
\clearpage
\section{Note 3: High temperature Brown-Zak oscillations}

\begin{figure}[h]
    \centering
    \includegraphics[width=1\linewidth]{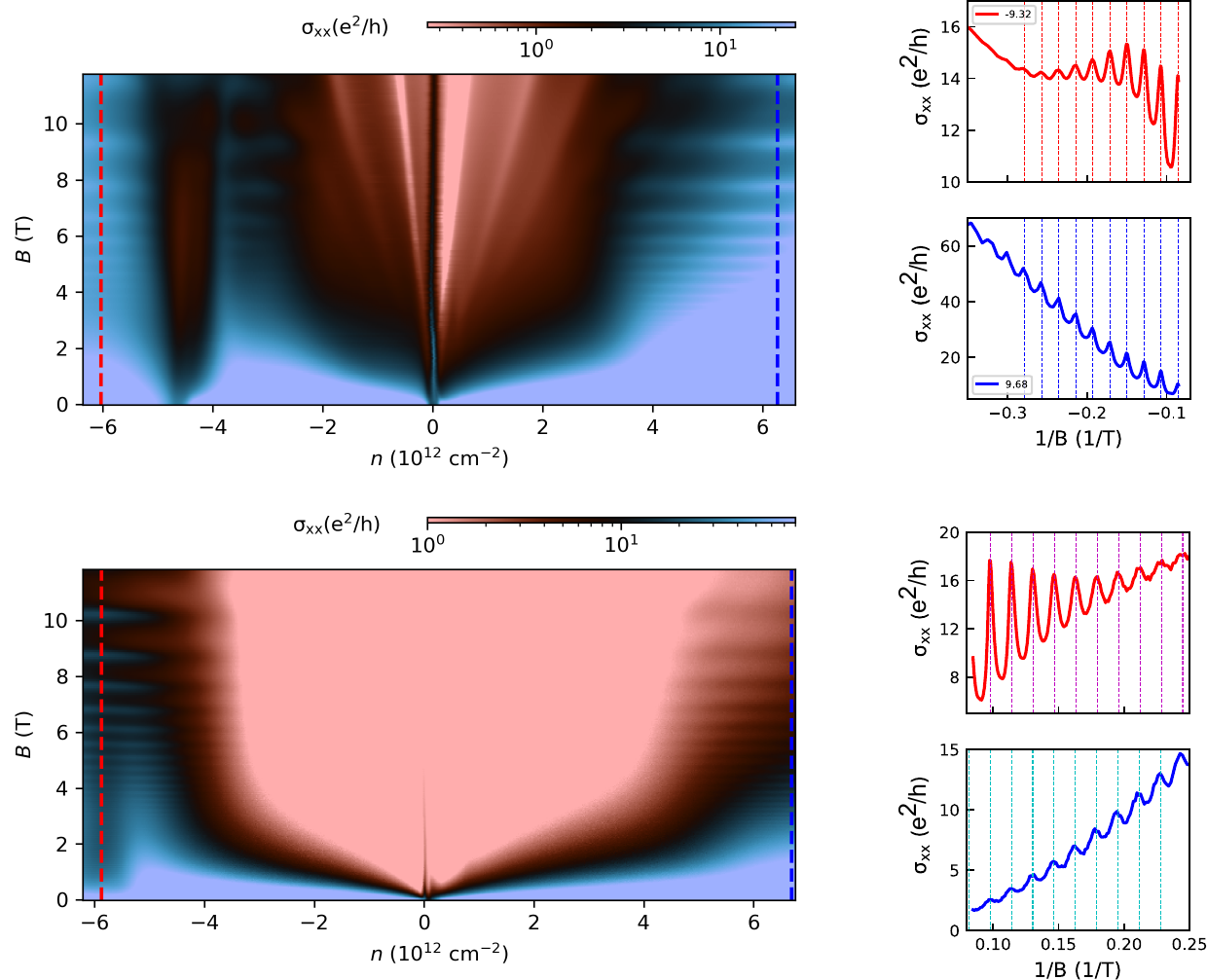}

    \caption{\textbf{Brown-Zak oscillations at high temperture.} Colormap of the resistance as a function of carrier density and high magnetic fields of the sample S2 (top) and S1 (bottom) at 110 \unit{\kelvin} when the top rotator is misaligned. In the right part, we show two cuts at the density shown in the red and blue dashed lines (respectively) showing the Brown-Zak oscillations.}
    \label{fig:Brown_Zak_B}
\end{figure}

In our samples the angular alignment with the bottom BN was extracted from high temperature magnetotransport measurements\cite{krishna_kumar_high-temperature_2017,wang_composite_2019}.
Figure \ref{fig:Brown_Zak_B}  shows the high temperature (110 \unit{\kelvin}) four probes longitudinal resistance as a function of magnetic field. The purpose of this is to  have access to the Brown-Zak oscillations, by removing the response coming from the quantum oscillations. From the periodicity of the Brown-Zak oscillations, we can extract the size of the bottom alignment. We show next to each Brown-Zak map the magneto-conductance at 110 \unit{\kelvin}, showing the Brown-Zak oscillations, that correspond to a fundamental field of $61.3$ \unit{\tesla} for samples S2, and 46.71 \unit{\tesla} for sample S1, corresponding to the same moiré lattice of $\lambda_{S2} = 10.11$ \unit{\nano\meter} and $\lambda_{S1} = 8.8$ \unit{\nano\meter}.

\section{Note 4: Super-moir\'e model}

\begin{figure}[h]
    \centering
    \includegraphics[width=0.35\linewidth]{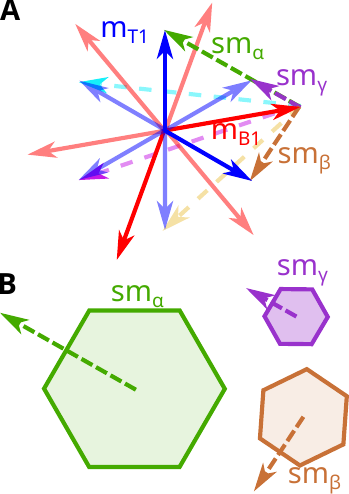}
    \caption{\textbf{Super-moir\'e model.} \textbf{A,} Reciprocal space vectors of the original moirés (solid line) in red and blue, and the first group of six super-moir\'e vectors (dashed lines). The alignments correspond to of $\theta_T = -0.59$\unit{\degree} and $\theta_B = 1.25$\unit{\degree}, respectively. The relative size of all the vectors is to scale. \textbf{B,} Corresponding area in reciprocal space of the three smallest composite moirés for this set of angles.}
    \label{fig:Composites}
\end{figure}

In the following we use the model previously developed in \cite{wang_composite_2019}. This model predicts the presence of mini-gaps at densities corresponding to the full filling of four electrons or 4 holes per area, where these areas (in reciprocal space) are given by linear combinations between the original moiré reciprocal lattice vectors, assuming that these areas are hexagonal. 
To give an example, let us consider the same set of angles as in the Fig. 2 of the main text. The original moirés' reciprocal lattice vectors are shown in red and blue for bottom (1.25\unit{\degree}) and top (-0.59\unit{\degree}), respectively in figure \ref{fig:Composites}. By subtracting them in the way shown in Fig.\ref{fig:Composites}A, we can obtain new vectors that define the areas in fig. \ref{fig:Composites}B. These areas are constructed as follows: the origin of the recently calculated vector serves as the center a regular hexagon whose apothem is half the length of this vector $G = \vert \vec{sm}_i \vert$, exactly as if we were constructing an hexagonal Brillouin zone. Then, the area is calculated as $A = \sqrt{3}G^2/2$, from where the density at full filling is $n=4\times A/(2\pi)^2 = \sqrt{3}G^2/2\pi^2$.
Notice that for the definition of these areas, the shape of the composite areas was imposed. In fact, this results in selecting only a subset of the possible new areas. This is because in order to define an area in 2 dimensions, 2 vectors are needed, and in here, only one was taken under the supposition that the area will keep the same shape of the original moirés (hexagonal), in a reminiscence of the definition of the Brillouin zone. However, nothing prevents us to define the areas using different combination of the original and new vectors, which ends up being in summary, a naive geometrical way to mathematically connect the composite model and the quasi-Brillouin zones model 
\cite{oka_fractal_2021}.

\begin{figure}[h]
      \includegraphics[width=.4\linewidth]{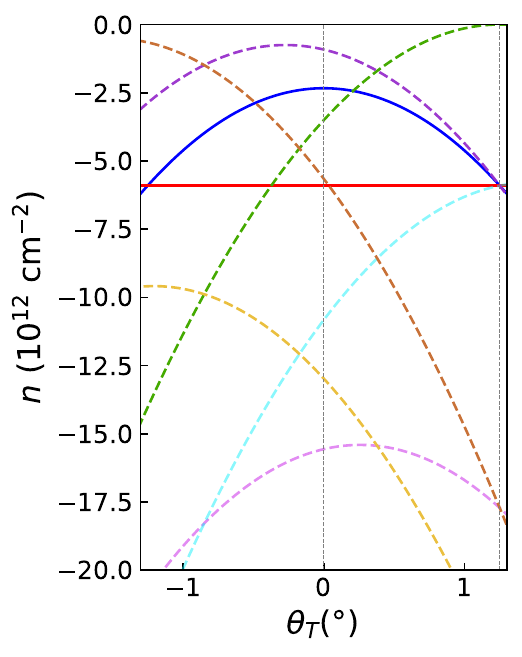}
     \includegraphics[width=0.45\linewidth]{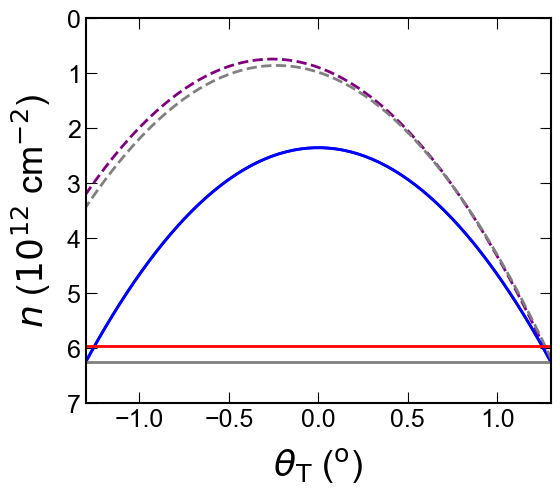}
    \caption{\textbf{(Left)}: Super-moiré model for $\theta_B =1.25$\unit{\degree}. This simulation corresponds to the sample in the main text. The red line shows the fixed bottom moiré, blue line corresponds to the moir\'e formed with the top rotator. Three of the six curves fall outside the range of densities studied in this article. \textbf{(Right)}: For the region in the experimental data, original moirés (solid - gray, red, blue) and super-moiré (dashed - gray, purple) calculated for two slightly different angles. The gray lines correspond to $\theta_{\mathrm{B}}= 1.30^\circ$, while the colored lines correspond to the sample S1 in the main text. We see that a shift bringing the bottom moiré to higher carrier densities also shifts the super-moiré to higher densities, contrary to what we see in figure 4 in the main text.}
    \label{fig:parabolas_high_dop}
\end{figure}

In Fig. \ref{fig:parabolas_high_dop}, we show the parabolas corresponding of all six first super-moir\'es represented in Fig. \ref{fig:Composites}A. Three of them fall outside the range of carrier density that can be reach in our experiments.
\subsection{The effective size of SM}
We mentioned in the text that the position of the SM peaks can give us the effective size of the super-moiré. This is because the super-moiré calculated with the model in \cite{wang_composite_2019} do not necessarily represent the size of the (quasi-) periodicity of the system. In fact, when calculating the commensurate approximant (see next section), the size that represents the system is generally larger than the super-moiré, see also \cite{oka_fractal_2021}. This makes that the densities at which we observe the SM peaks are reached at a doping of $n\times 4$ electrons/holes per area of the commensurate approximant. For example, when considering the angles $\theta_B = 1.25^\circ$ and $\theta_T = -0.12^\circ$, the effective super-moiré size (purple) $\lambda_{SM} \simeq 21.3$ \unit{\nano\meter}, however the commensurate approximant is $\lambda_{CA} \simeq 64$ \unit{\nano\meter}. In this case, the SM peak is located at a density of $9\times 4$ electrons/holes per area of the commensurate approximant. 

When considering such a large periodicity, many nmini-gaps are ideally expected to appear in the electronic structure, however as the super-moiré interaction is not strong enough, several mini-gaps can not be observed. The most pronounced peaks are the ones satisfying the super-moiré model in \cite{wang_composite_2019} or in the previous section.

\begin{figure}[h]
    \centering   
      \includegraphics[width=.6\linewidth]{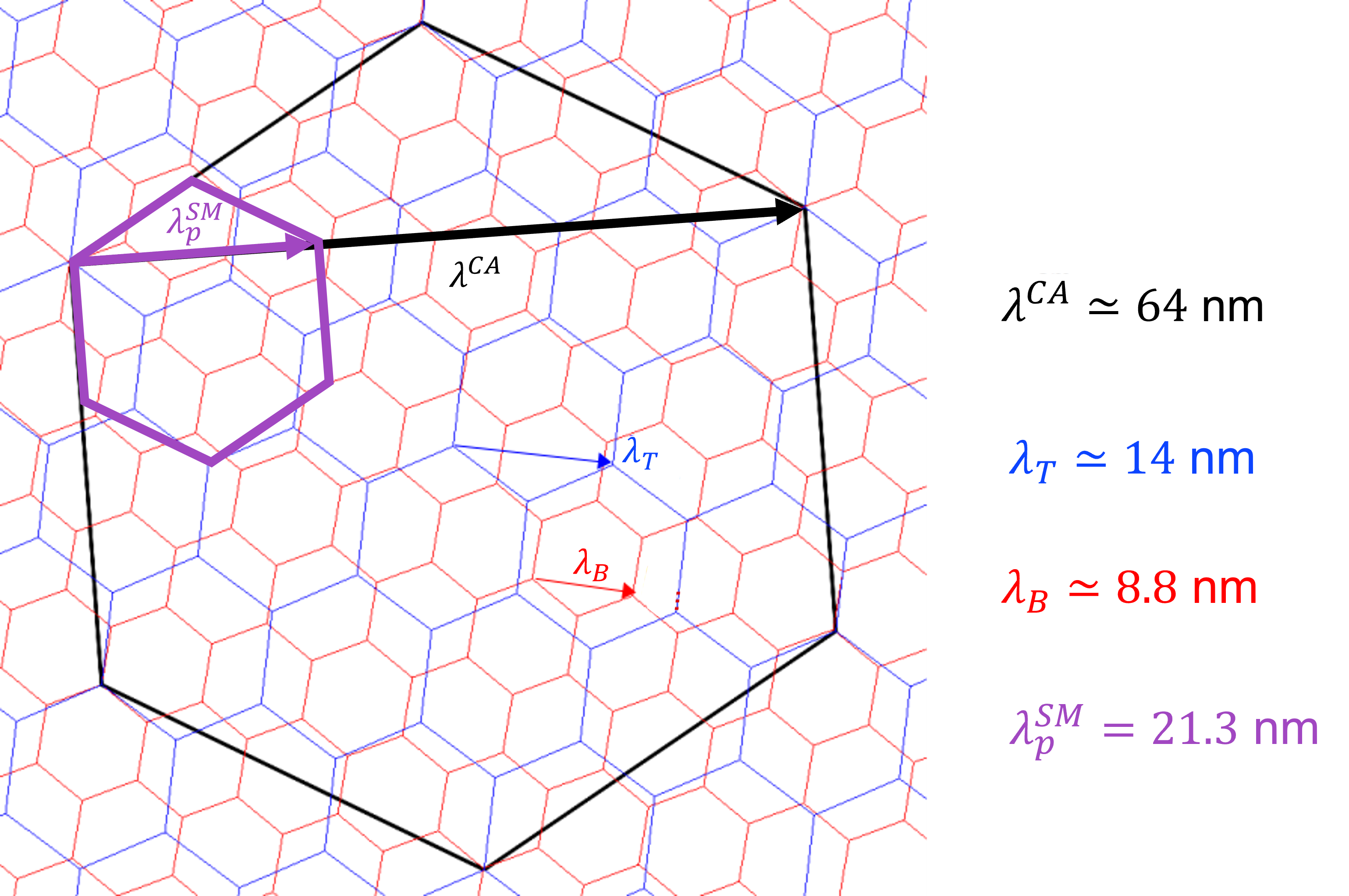}
    \caption{\textbf{Comparison of sizes of commensurate approximant and super-moiré}. Schematics of the sizes of the moirés and super-moirés formed for angles $\theta_B = 1.25^\circ$ and $\theta_B = -0.12^\circ$}
    \label{fig:sizes}
\end{figure}

\clearpage
\section{Note 5: Simulation: relaxation and commensurate approximants}
The structural relaxation is obtained using the LAMMPS molecular dynamics (MD) simulation package \cite{thompson_LAMMPS_2022, plimpton_fast_1995} for which we combine parametrized force-fields \cite{brenner_second-generation_2002,kinaci_thermal_2012,leven_interlayer_2016} for modeling the considered graphene/hBN heterostructures. In particular, the second-generation REBO potential \cite{brenner_second-generation_2002} is used to compute the couplings of carbon atoms while the Tersoff potentials \cite{kinaci_thermal_2012} are for B-N interactions and the ILP ones \cite{leven_interlayer_2016} for the interlayer couplings between graphene and hBN layers. In all simulations, the in-plane lattice mismatch between graphene and hBN ($\sim 1.78\%$) was taken into account.

In addition, hBN/graphene/hBN heterostructures are generally incommensurate, i.e., there is no real periodic cell. To perform the above-mentioned relaxation calculations, we therefore have to use commensurate approximants \cite{oka_fractal_2021}. This approximation is based on the fact that in any case ($\theta_B$,$\theta_T$), there always are pairs of lattice points of the two moiré patterns which happen to be very close to each other. This is described as
\begin{eqnarray}
    n_1^T \textbf{L}_1^T + n_2^T \textbf{L}_2^T &=& n_1^B \textbf{L}_1^B + n_2^B \textbf{L}_2^B+\Delta \textbf{L}.
\end{eqnarray}
where $n_i^{T,B}$ are integers and $\mathbf{L}^{T,B}_{1,2}$ are the supperlattice vectors of top and bottom moirés. When $\left| \Delta \textbf{L} \right| \ll \left| n_1^T \textbf{L}_1^T + n_2^T \textbf{L}_2^T \right|$ and $\left| n_1^B \textbf{L}_1^B + n_2^B \textbf{L}_2^B \right|$ (see an example in Fig.S6), the commensurate superlattice obtained by neglecting $\Delta \textbf{L}$ can be a good approximant for modeling the considered super-moiré system. Basically, good commensurate approximants can always be obtained with the reasonably large integers $n_i^{T,B}$ (see similar discussions in \cite{oka_fractal_2021}).

\begin{figure}[h]
    \centering
    \includegraphics[width=1\linewidth]{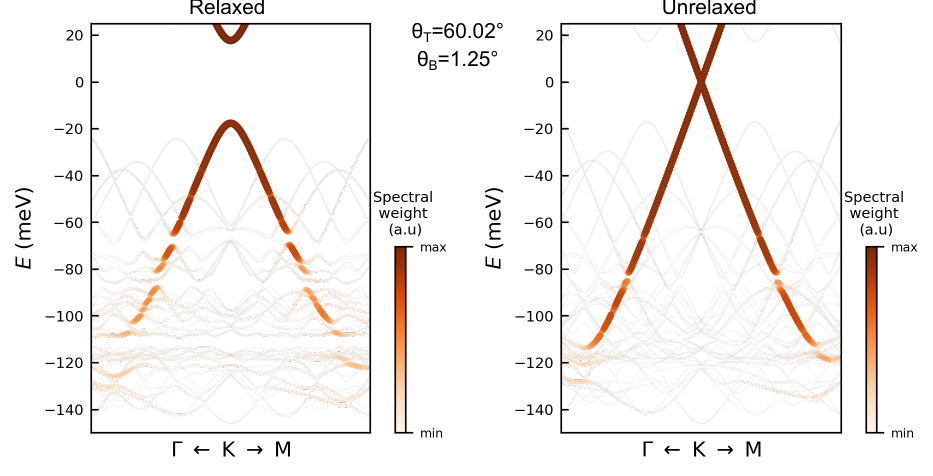}
    \caption{Band structure of the double moiré system of $60.02^\circ$ and $1.25^\circ$ for the relaxed and unrelaxed structures showing differences in the amplitudes of the gaps.}
    \label{fig:RelUnrel}
\end{figure}

\subsection{Simulation: electronic structure}

To compute the electronic structure of the considered graphene/hBN systems, we employed the $p_z$ tight-binding Hamiltonian, which is in the similar form as presented in \cite{trambly_de_laissardiere_localization_2010, moon_electronic_2014}. In particular, the Hamiltonian is written as
\begin{equation}
    H_{tb} = \sum_{n} V_n a^\dag_n a_n + \sum_{n,m} t_{nm} a^\dag_n a_m \nonumber
\end{equation}
where the on-site energies $V_n$ = 0, 3.34 eV, and -1.4 eV for carbon, boron, and nitride atoms, respectively. The hopping energies $t_{nm}$ are determined using the standard Slater-Koster formula
\begin{eqnarray}
	t_{nm} (r_{nm}) &=& V_{pp\pi} \sin^2 \phi_{nm} + V_{pp\sigma} \cos^2 \phi_{nm} ,  \nonumber \\
	V_{pp\pi} &=& V^0_{pp\pi} \exp \left( (a_0 - r_{nm})/r_0 \right), \nonumber \\
	V_{pp\sigma} &=& V^0_{pp\sigma} \exp \left( (d_0 - r_{nm})/r_0 \right)  \nonumber	
\end{eqnarray}
where the direction cosine of ${\vec r}_{nm}$ along Oz axis is $\cos \phi_{nm} = z_{nm}/r_{nm}$, $r_0 = 0.184a$, $a_0 = a/\sqrt 3$, and $d_0 = 3.415$\AA ~ while $a \simeq 2.46$\AA.

\begin{figure}[h]
    \centering
    \includegraphics[width=.75\linewidth]{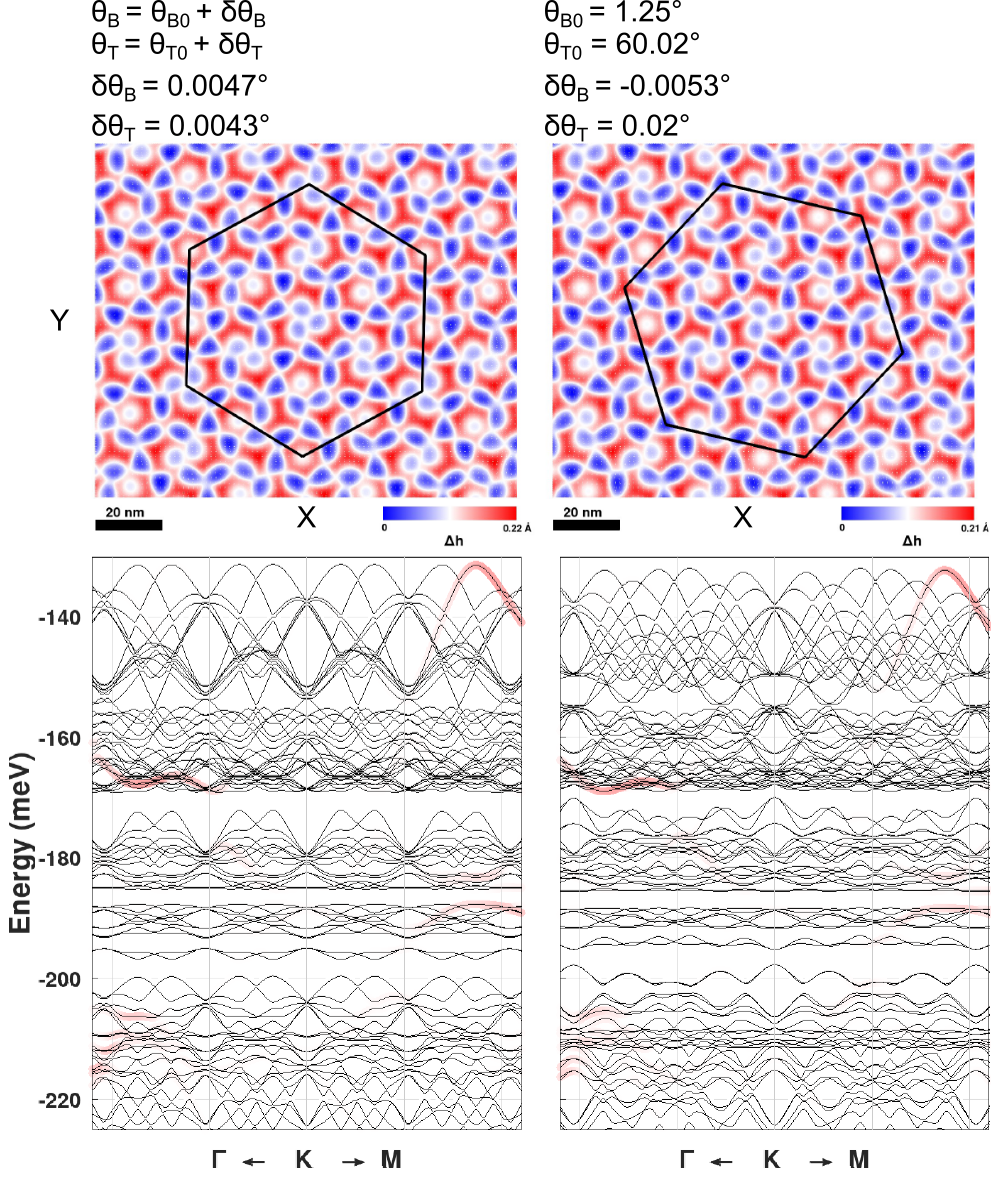}
    \caption{Buckling pattern and band structure for 2 different commensurate approximants, obtained by slightly changing the top and bottom moiré angles. The changes in band structure are only noticeable for the high doping regime.}
    \label{fig:CommAppDiff}
\end{figure}

\clearpage
\section{Note 6: Measurements in samples with different bottom BN}

In the following we show data of sample S2 and S3. Sample S2 has the same BN but different alignment than S1. Sample S3 has a different BN.

\begin{figure}[h]
    \centering
    \includegraphics[width=.33\linewidth]{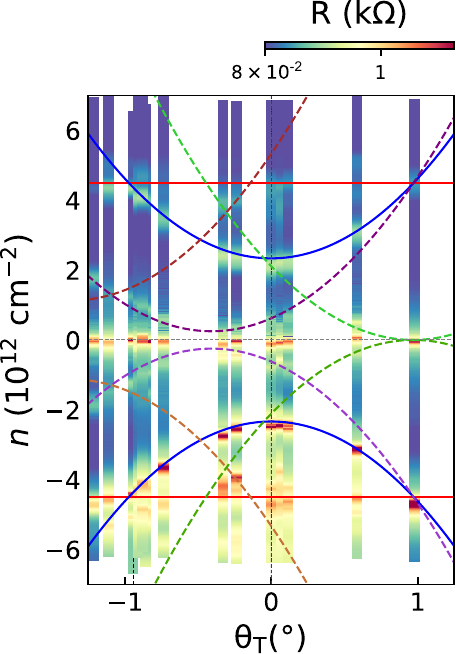}
    \caption{Colormap of the resistance as a function of density for the different angles measured for sample S2 with a bottom moiré of $10.11$ \unit{\nano\meter} or $0.97$\unit{\degree}.}
    \label{fig:map}
\end{figure}

\begin{figure}[h]
    \centering
    \includegraphics[width=.34\linewidth]{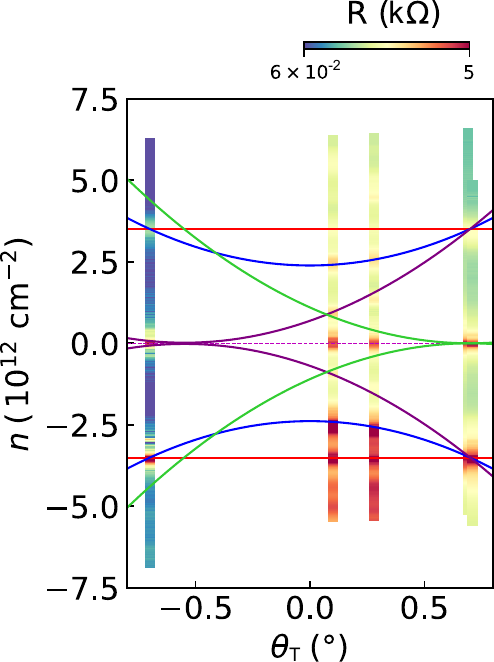}
    \caption{Colormap of the resistance as a function of density for the different angles measured for sample S3 with a bottom moiré of $11.6$ \unit{\nano\meter} or $0.7$\unit{\degree}. For the angles at -0.7\unit{\degree}, 0.1\unit{\degree} and 0.3\unit{\degree}, features that can be associated with qBZ are clearly visible for densities higher than the red line (position of the bottom moiré). The reduced number of points prevent us to do a proper identification of these qBZ.}
    \label{fig:LSFM1}
\end{figure}

\clearpage
\subsection{qBZ in sample S2}

We show the mini-gaps as a function of density and angular alignment for sample S2. We superposed the single moiré, super-moiré and qBZ curves, similar to the Fig.3 in the main text. We also give the list of the combinations of numbers $(p,q,r,s)$.

\begin{figure}[h]
    \centering
    \includegraphics[width=1\linewidth]{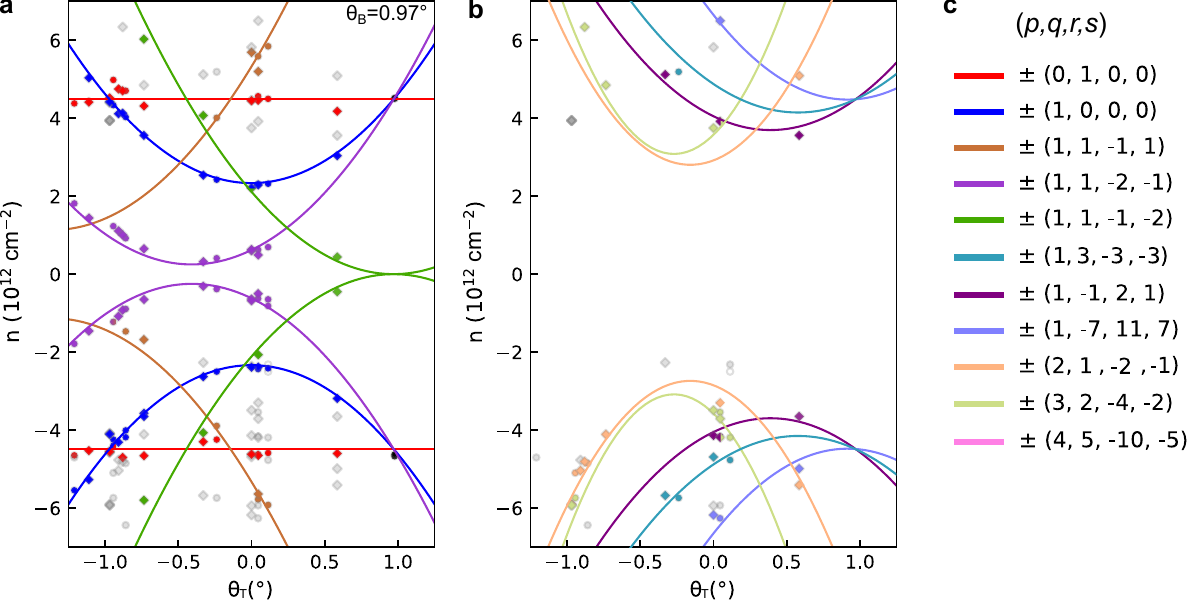}
    \caption{Main gaps and qBZ for the sample S2 with a fixed moiré of $10.11$ \unit{\nano\meter} or $0.97$\unit{\degree}. This plot is similar to figure 3 in main paper. In this sample, we can observe a similar behavior than the sample in the main text, but with different sets of $(p,q,r,s)$ because the bottom angle is different.}
    \label{fig:JV3A_points}
\end{figure}

\newpage
\section{Note 7: Coefficients equivalence with 2CN}

As mentioned in the last part of the main text, we can calculate the 6 coefficients $\{v_{ij}\}$ from the original 4 coefficients by following \cite{koshino_topological_2022}. There is a simple relationship to calculate $\{v_{ij}\}$ from the $\{p,q,r,s\}$ numbers. We resume the numbers in the following equation:
\begin{equation}
\begin{aligned} 
    \{v_{ij}\} = \{&p, (2r-s)/3, (r+s)/3,\\
    & (r-2s)/3, (2r-s)/3, q\}
\end{aligned}
\end{equation}
To obtain the second Chern numbers, we should apply the relation:
\begin{equation}
    \{C_{ij}\}= -\{v_{ij}\}. 
\end{equation}

To sum up the obtained numbers, we show the parabolas in figure 3 with the original $\{p,q,r,s\}$ and the corresponding $\{v_{ij}\}$.

\begin{figure}[h!]
    \centering
    \includegraphics[width=1\linewidth]{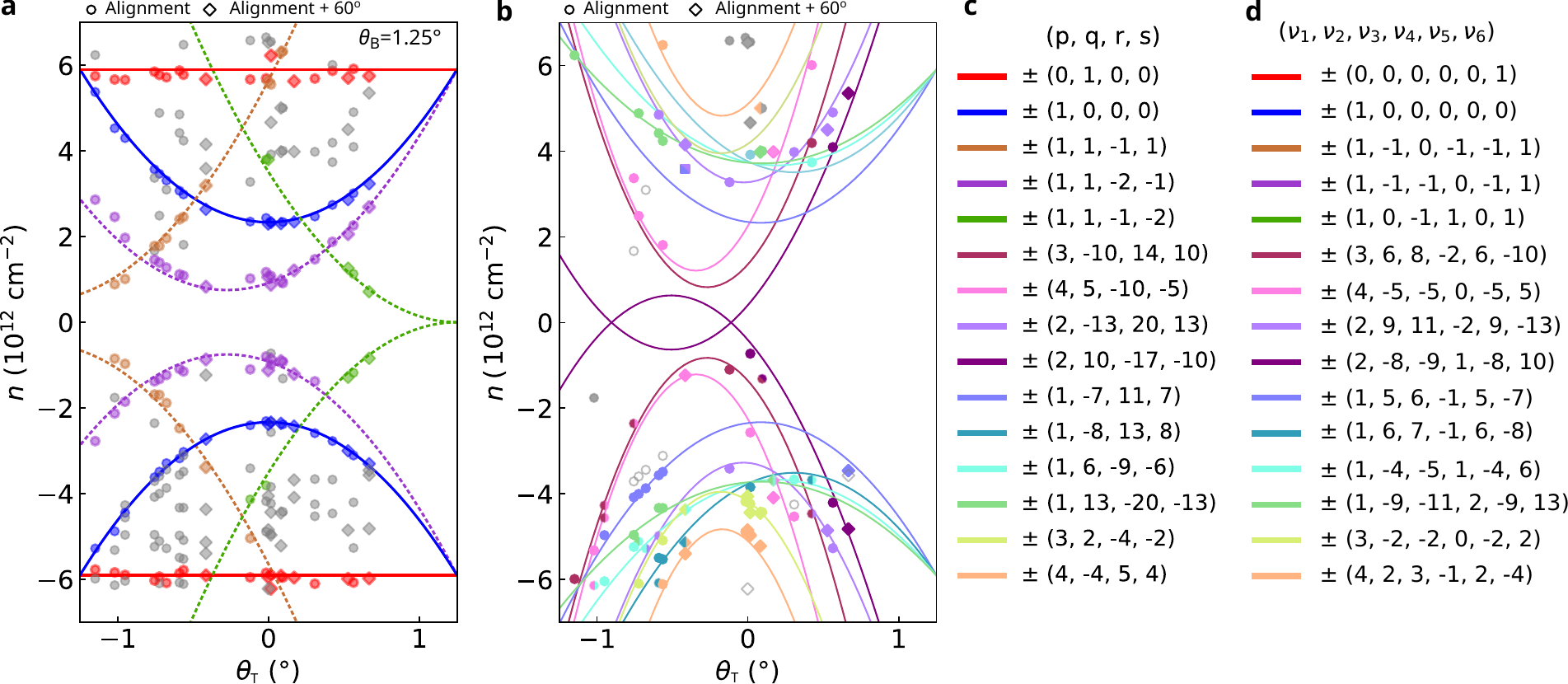}
    \caption{Same figure as the fig. 3 in the main text, resuming the mini-gaps as a function of carrier density and angular alignment, superposed with the calculated qBZ. \textbf{d} Corresponding $\{v_{ij}\}$ for each set $(p,q,r,s)$. $\{v_{ij}\}= -\{C_{ij}\}$ allows for the calculation of the second Chern numbers.}
    \label{fig:2CN}
\end{figure}

\clearpage
\section{Note 8: Effects of different rotation center in the band structure}
\begin{figure}[h!]
    \centering
    \includegraphics[width=0.85\linewidth]{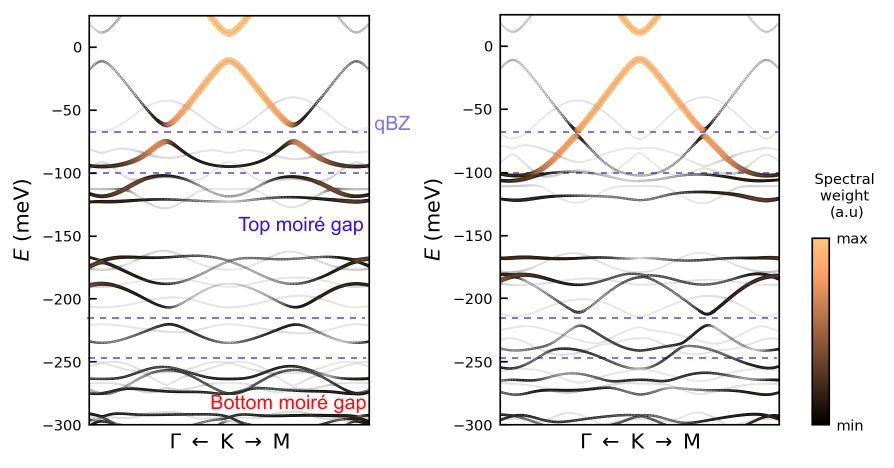}
    \caption{Full energy range of the band structures calculated in Fig 4 in the main text. Only the qBZ gaps are affected by the change in the rotation center, and the original moiré a}
    \label{fig:full_range}
\end{figure}

\bibliography{references_RR}